# A Voice-Enabled Virtual Patient System for Interactive Training in Standardized Clinical Assessment

Veronica Bossio-Botero[1,2], Vijay Yadav[1,3], Jacob Ouyang[1], Anzar Abbas[1], Michelle Worthington[1,4]

[1]Brooklyn Health, Brooklyn, NY, United States
[2]Columbia University, Zuckerman Institute, New York, NY, United States
[3]School of Psychology, University of New South Wales, Sydney, Australia
[4]Yale University School of Medicine, Department of Psychiatry, New Haven, CT, United States

## Abstract

**Background:** Training mental health clinicians to conduct standardized clinical assessments is challenging due to a lack of scalable, realistic practice opportunities. Traditional methods often fail to prepare trainees for the variability and complexity of real-world patient interactions, potentially impacting data quality in clinical trials. This paper introduces a novel approach to address this training gap using large language model (LLM)-based interview simulations.

**Objective:** This study aims to develop and validate a voice-enabled virtual patient simulation system as a proof-of-concept. We describe the development of the system and evaluate whether it can generate virtual patients that (1) accurately adhere to pre-defined clinical profiles, (2) maintain a coherent and consistent narrative, and (3) produce dialogue that is perceived as realistic.

**Methods:** We implemented a system that uses a LLM to simulate patients with specified symptom profiles, demographic backgrounds, and distinct communication styles. The system's performance was analyzed through a mixed-methods evaluation, which included a formal assessment by 5 experienced clinical raters who conducted simulated structured MADRS interviews on 4 virtual patient personas, scored them on the scale and provided qualitative feedback on the system's clinical plausibility, narrative cohesion and dialogue realism.

**Results:** Across a total 20 interviews, the virtual patients demonstrated strong adherence to their configured clinical profiles, with human raters scoring the patients with high accuracy against their predefined score configurations. The mean item difference between MADRS rater scores and configured scores was 0.52 (SD = 0.75); Inter-rater reliability across items was 0.90 (95% CI =0.68-0.99). Expert raters consistently gave average ratings of "Agree" to "Strongly Agree" when asked to evaluate the qualitative realism and cohesion of the virtual patients.

**Conclusions:** LLM-powered virtual patient simulations promise a viable and scalable new tool for training clinicians in standardized clinical assessment. This pilot study demonstrates the system's ability to produce high-fidelity, clinically relevant practice scenarios.

## Introduction

The clinical interview is a fundamental tool in psychiatric assessment and serves as a primary endpoint in clinical trials for central nervous system (CNS) disorders. Because most psychiatric conditions lack objective biomarkers, the field relies on subjective clinical evaluations. The validity of clinical trials, and consequently, the development of new therapeutics, thus fundamentally depends on the quality of these assessments. Despite its critical role, the development of consistent interviewing skills remains a significant challenge, as conventional training methods frequently fail to provide the scalable and realistic practice environments for this complex task.

High inter-rater reliability in clinical assessments is essential for minimizing data noise and identifying the true effects of treatment. However, extensive evidence indicates that conventional clinical rater training approaches, such as peer role-play, are insufficient for achieving this standard, showing little to no improvement in the validity of trainee assessments [1-3]. Even when initial training is successful, rater scores often shift over the course of a trial, a phenomenon known as “rater drift”, which can further compromise data quality [2,4-6]. At the core of this issue is a fundamental methodological gap: the lack of scalable, on-demand, and standardized “patient stimuli” to calibrate and maintain rater performance over time.

Current methods for clinical interview training include watching pre-recorded videos or conducting role-play with peers, both of which fail to replicate the realism and interactive nature of real clinical scenarios. The certification process for clinical trial raters introduces a further systemic obstacle. Sponsors typically require raters to meet strict criteria, often including a minimum number of completed scale administrations. Because uncertified raters cannot access opportunities to conduct assessments, they cannot meet the experience thresholds required for certification. This circular barrier limits the pool of trained raters and perpetuates poor measurement practices.

The use of simulation to bridge the gap between training and clinical practice has been an important tool in medical education for decades, with applications in many specialties [7-10]. Standardized Patients (SPs)—individuals trained to consistently portray specific clinical scenarios—provide a more realistic alternative to peer role-play for teaching communication and diagnostic skills. However, they have significant limitations as they are expensive and difficult to scale, which limits the repeated practice opportunities that are essential for skill development [11]. Moreover, the human variability in SP performance can compromise the very standardization they aim to achieve [12,13].

To address these limitations, the field has developed successive generations of virtual patients—computer systems designed to simulate clinical scenarios. While virtual patients evolved from simple branching narratives to being capable of processing free-text input, their reliance on matching user inputs to a fixed database of pre-scripted responses was a fundamental limitation [14-17]. Consequently, the trade-off between the scalability of automated systems and the realism of SPs has persisted.

Large language models (LLMs) offer a solution to this long-standing challenge by shifting the mechanism of simulation from information retrieval to generation of novel content. By synthesizing context-aware language in real-time, these models are able to simulate standardized clinical scenarios while maintaining flexibility and realism. Emerging empirical evidence supports this utility: randomized trials show that LLM-based simulations achieve high dialogue authenticity ratings and reduce trainee anxiety compared to live observation, providing an accessible environment for skill development [18-22]. As Zeng and colleagues note in a 2025 scoping review [23], while LLM-based virtual patients are still in early stages of development, they hold significant potential for improving medical training.

Here, we describe the development and initial evaluation of a voice-enabled virtual patient system designed specifically for CNS clinical trials. The system uses a full speech-to-speech pipeline to simulate highly interactive and realistic clinical assessments. As an initial evaluation of the system, we tested the alignment of expert-rated symptom severity scores on the Montgomery-Asberg Depression Rating Scale (MADRS) [24] with pre-defined scores for virtual patient personas. The primary objectives of this pilot study are to determine if the system can generate virtual patients that: (1) adhere to pre-configured clinical profiles with sufficient accuracy for expert raters to score them reliably (2) maintain a coherent and consistent narrative throughout an interview; and (3) produce dialogue that clinical experts perceive as realistic.

# Methods

## System design

Our system provides a simulated clinical encounter where a user can practice conducting an interview-based clinical assessment on a virtual patient using a speech-to-speech pipeline. The system's architecture (Figure 1) is composed of two primary components: a patient profile engine for constructing clinically plausible and narratively coherent personas dynamically, and an interactive voice pipeline that animates these personas through real-time, spoken dialogue.

*Patient Profile Generation*

The core of each profile is a randomly generated set of scores for the symptoms defined by the MADRS, which collectively define the clinical presentation. Importantly, this random generation was constrained by logical consistency checks to ensure the co-occurrence of symptoms is clinically plausible and representative of actual presentations [25]. For example, a high score for suicidal ideation must be accompanied by a high score for depressed mood.

In addition to the clinical profile, each virtual patient is assigned a specific communication style that modulates how their symptoms are communicated. The styles are: cooperative, a baseline style where responses are designed to provide clear and scorable information and guarded, which reflects the communication barriers often present in depression where a patient may be reluctant to disclose details. Finally, each profile is contextualized within a unique demographic and situational narrative, including a name, age, and personal history.

The target item scores and their corresponding behavioral descriptions are combined with the demographic information and the communication style into a detailed system prompt. This prompt serves as the complete specification for the virtual patient's persona and is provided to the core large language model (Anthropic's Claude Sonnet 3.7 [26]) to guide its dialogue generation throughout the interview, ensuring each interaction is grounded in a consistent and clinically valid profile.

*Interactive Voice Pipeline*

The interactive dialogue is driven by a real-time speech-to-speech pipeline. The loop begins with the clinician's spoken input, which is captured and transcribed using Amazon Transcribe's streaming speech-to-text (STT) service [27]. The resulting text is then passed to the language model, which generates a response based on the patient profile and the evolving conversation history.

This response is then converted back into speech using a text-to-speech (TTS) model. The system supports several high-fidelity TTS providers—including Amazon Polly, ElevenLabs, and Cartesia—allowing for flexible control over vocal characteristics and degrees of realism. The interview continues until the clinician determines they have sufficient information for a clinical assessment, at which point their scores, qualitative feedback, and the interview transcript are recorded.

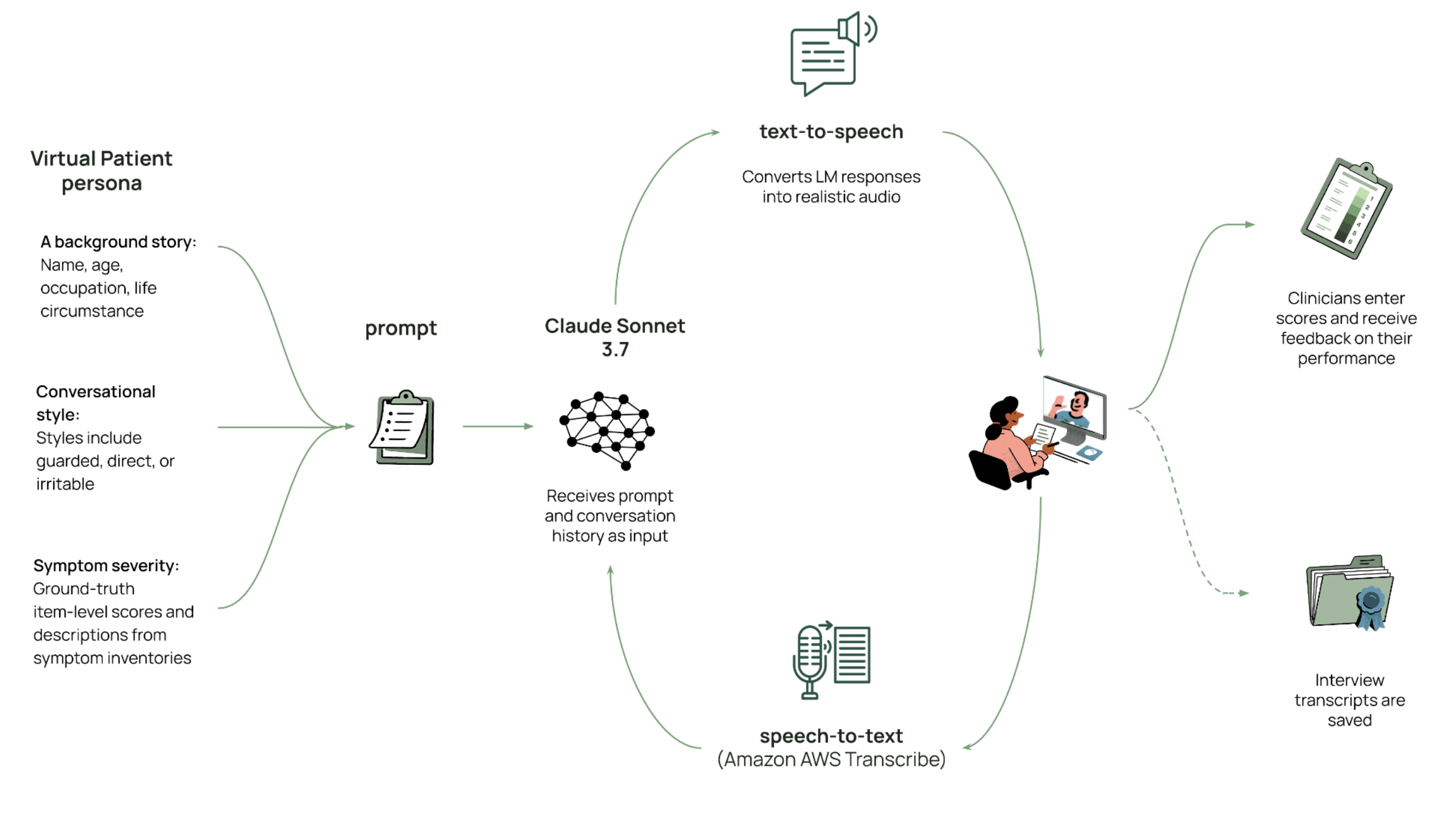

**Figure 1.** System architecture and interaction workflow. The clinician's speech is transcribed by a speech-to-text (STT) module. The resulting text is processed by an LLM, which generates a response grounded in a detailed system prompt (defined by the patient's symptoms, background, and style). A text-to-speech (TTS) model synthesizes the LLM's response into audio, completing the real-time, turn-based dialogue loop.

## User study

To evaluate the system's performance and perceived quality, we conducted a user study with trained clinicians who had extensive experience administering the MADRS. The study was designed to assess the virtual patients' adherence to their clinical profiles by testing the alignment of the rater-assigned MADRS scores with the pre-defined MADRS scores and to gather expert feedback on the realism and coherence of the simulated dialogue. This study was reviewed by the Biomedical Research Alliance of New York (BRANY) Review Board and was determined to be exempt from review (protocol #25-196-2265).

*Participants*

A group of expert raters (N=5) was recruited through our professional network. Participants were required to have a master's degree or higher in psychology, psychiatry, social work or a related field and have prior experience administering the MADRS.

*Materials*

We created a fixed set of four virtual patient personas for the study. The personas were balanced for gender (two male, two female) and communication style (two cooperative, two guarded). The underlying preconfigured profiles were designed to span a wide range of symptom severities as defined by the MADRS, with total scores ranging from 15 (mild) to 39 (severe). All expert raters interacted with the same four virtual patients to ensure consistency and allow for inter-rater reliability analysis.

The core dialogue engine was driven by Anthropic's Claude Sonnet 3.7 large language model [26]. Speech-to-text (STT) transcription was handled by Amazon Transcribe [27], and speech synthesis (TTS) was generated using ElevenLabs voice models [28].

*Procedure*

The study was conducted remotely via video call. After providing informed consent, each participant received a brief onboarding to demonstrate the system's voice-based interface and to review the study tasks. Participants were instructed to administer the MADRS for each virtual patient, following the Structured Interview Guide for the MADRS (SIGMA) methodology [29].

For each of the four personas, the procedure was as follows:

1. The expert rater conducted the full voice-based interview with the virtual patient using the speech-to-speech system.
2. Immediately following the interview, the rater was directed to a post-interview survey using a link where de-identified responses were collected.
3. In the survey, the rater first provided their clinical scores for the virtual patient on each of the 10 MADRS items.
4. After submitting their scores, the rater was presented with a comparison chart showing their ratings alongside the virtual patient's pre-programmed profile to inform the rater's assessment of adherence to clinical profile. They were then asked to complete the qualitative feedback portion of the survey.

*Measures*

**Clinical scoring:** Raters scored the virtual patient on each of the 10 items of the MADRS based on the evidence gathered during the interview.

**Qualitative ratings:** Raters indicated their agreement on a 7-point Likert scale (1-Strongly Disagree to 5-Strongly Agree) with three statements assessing:

- Profile consistency: “The behavior of the virtual patient during the interview was consistent with the symptom profile it was configured to exhibit.”

  As this question was presented after raters saw the pre-configured scores, it was not designed to re-evaluate scoring alignment (which was captured by the quantitative “clinical scoring”), but rather as a meta-assessment of portrayal quality. Its purpose was to determine whether any observed discrepancies between the expert’s initial score and the simulation configuration were still perceived as falling within a clinically plausible range.

- Dialogue realism: “The dialogue produced by the virtual patient was realistic and natural in terms of language structure and vocabulary.”
- Character cohesion: “The virtual patient maintained a coherent character throughout the interview, without factual or behavioral contradictions.”

The qualitative part of the survey also included space for optional open-ended comments on the Likert ratings.

## Results

### Alignment of pre-defined and rater-scored MADRS scores

Across all MADRS items, which are each rated on a 0-6 scale, the mean difference between rater and preconfigured scores was 0.52 (SD = 0.75). The distributions of these item-level score differences were clustered around zero but exhibited a consistent rightward skew (Figure 2), indicating that raters tend to assign slightly higher scores than the target scores. This trend was stable across all four personas, with mean item score differences of 0.60, 0.60, 0.66 and 0.46 respectively (SD of means = 0.07) (Figure 3).

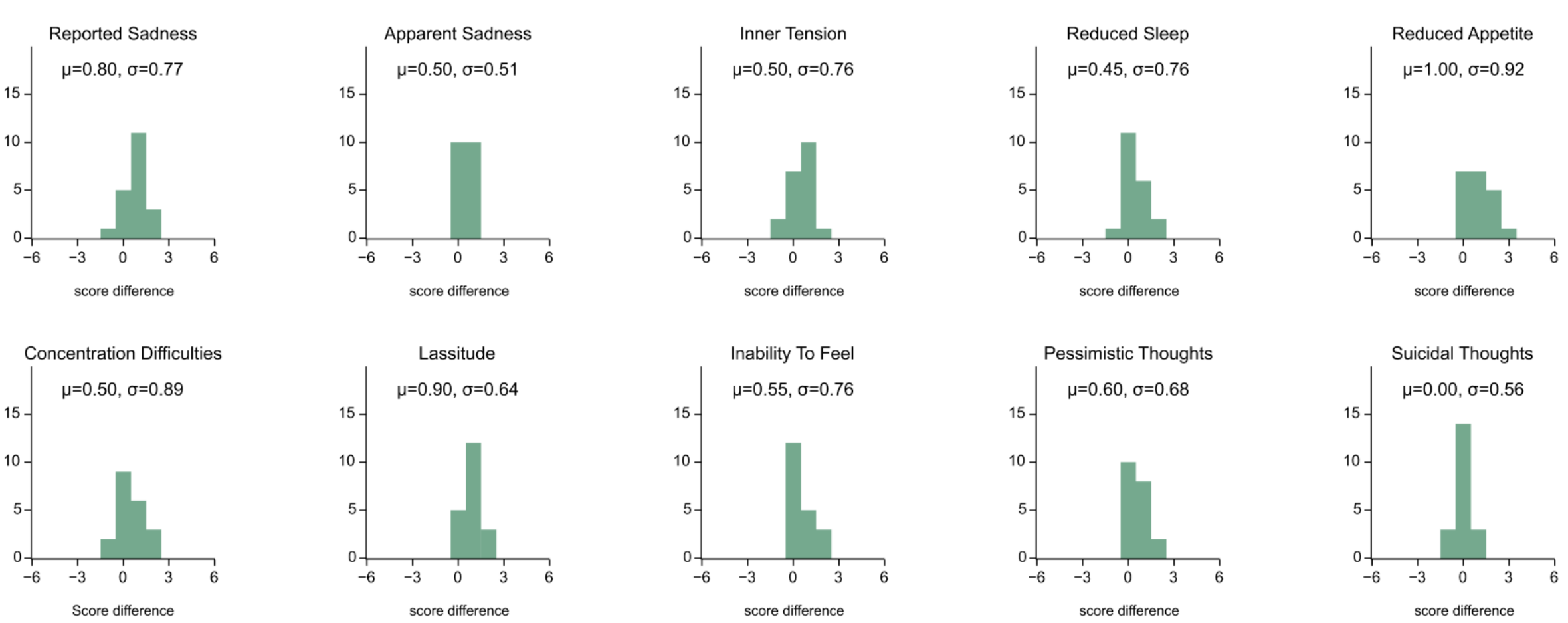

**Figure 2.** Histograms illustrate the distribution of score differences (clinician score minus pre-configured score) for each item of the MADRS.

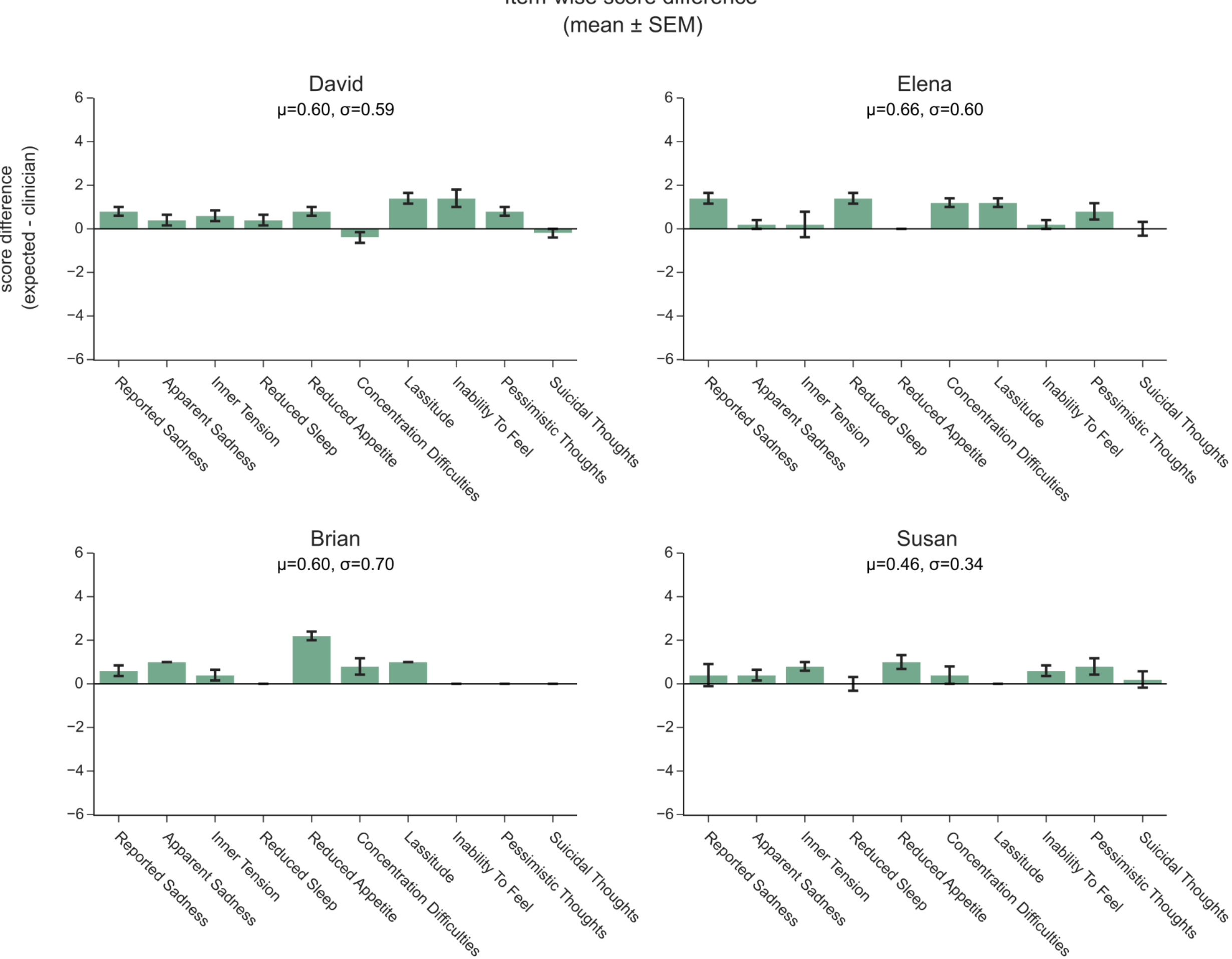

**Figure 3.** Average score differences between clinician score and pre-configured score for each MADRS item, shown for the four patient personas. Error bars represent the standard error of the mean across raters.

The cumulative effect of these small but systematic overestimations at the item level resulted in a consistent difference in the total MADRS score (the overall sum of the item-level scores). While the mean total MADRS scores assigned by raters tracked the intended severity across all personas (Figure 4), clinicians scores averaged moderately higher than the configured ground truths (mean absolute error = 5.80 ± 0.73 s.d.).

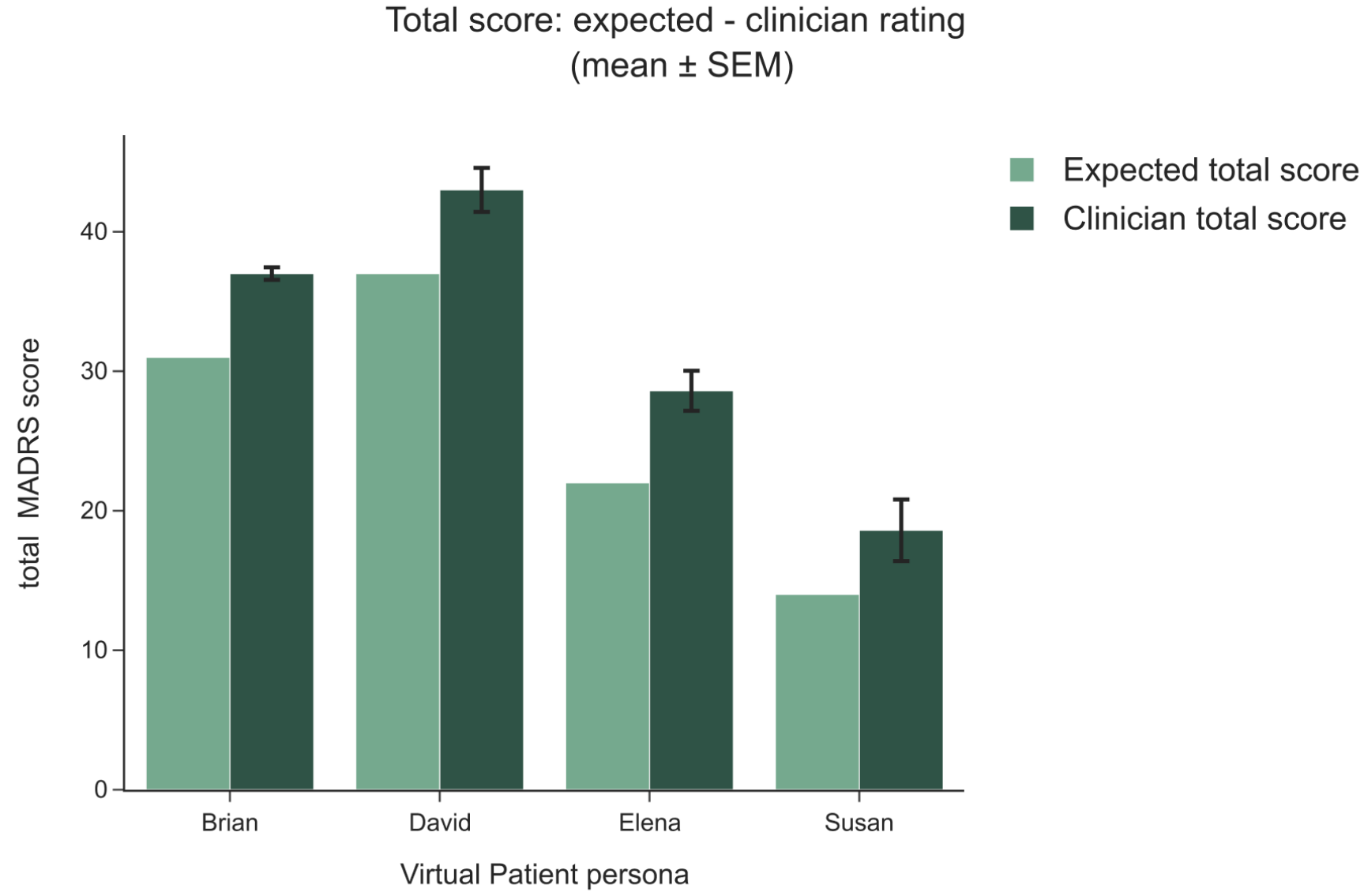

**Figure 4.** Total MADRS score from the virtual patient configuration (light green) versus total MADRS score assigned by the clinicians (dark green) for each of the four patient personas.

## Inter-rater reliability across expert raters

The virtual patients elicited consistent ratings across the raters in the study. Inter-rater reliability for the total MADRS score was high, with an intraclass correlation coefficient (ICC) of 0.90 based on a two-way random effects model for a single rater's absolute agreement (ICC(2,1)). Reliability for individual items was variable, ranging from excellent for items such as 'Lassitude' (ICC = 0.98) and 'Reduced Appetite' (ICC = 0.96) to fair for 'Inner Tension' (ICC = 0.63) and 'Concentration Difficulties' (ICC = 0.69).

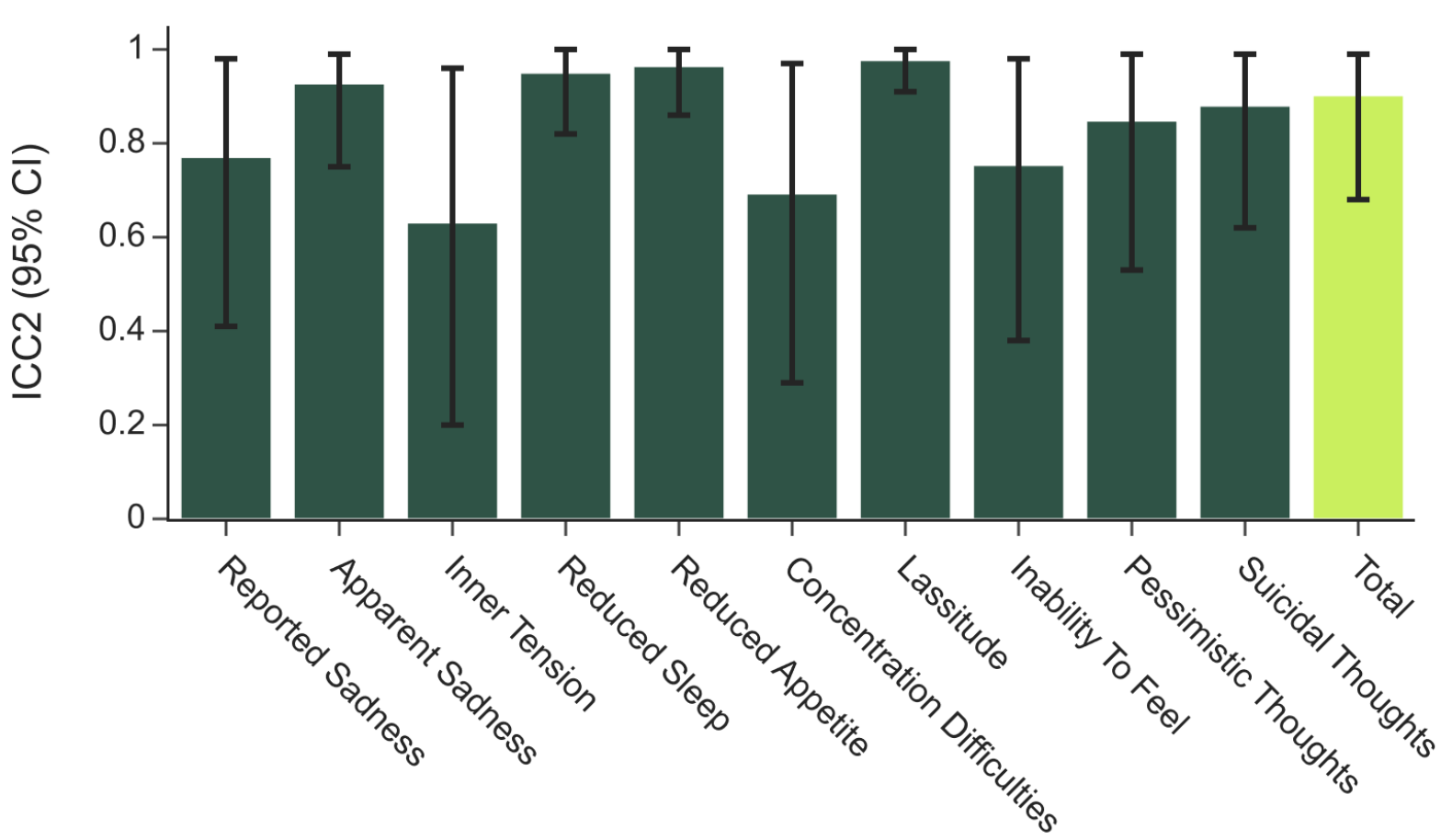

**Figure 5.** Intraclass correlation coefficients between raters for each item score (dark green) and for the total score (bright green) of the MADRS. Error bars represent 95% confidence intervals.

### Assessing profile consistency, coherence, and realism

Qualitative ratings for the 20 interviews indicated broadly favorable perceptions of the system's performance. When asked if the patient's behavior was consistent with its configured profile, the ratings for all interviews were positive, including 'Strongly Agree' (n=5), 'Agree' (n=14), and 'Somewhat Agree' (n=1). In 19 out of the 20 interviews, the character cohesion was judged favorably, with ratings to the statement "The virtual patient maintained a coherent character throughout the interview, without factual or behavioral contradictions" including 'Strongly Agree' (n=11), 'Agree' (n=4), and 'Somewhat Agree' (n=4). One rating was 'Somewhat Disagree' (n=1), with the rater commenting that the virtual patient was factually consistent but at times behaviorally inconsistent. Ratings for linguistic naturalness were also high, with 19 of 20 responses being positive: 'Strongly Agree' (n=9), 'Agree' (n=9), and 'Somewhat Agree' (n=1). One interview received a rating of 'Somewhat Disagree' (n=1), for which the rater commented that the virtual patient was unrealistically verbose compared to a severely depressed person.

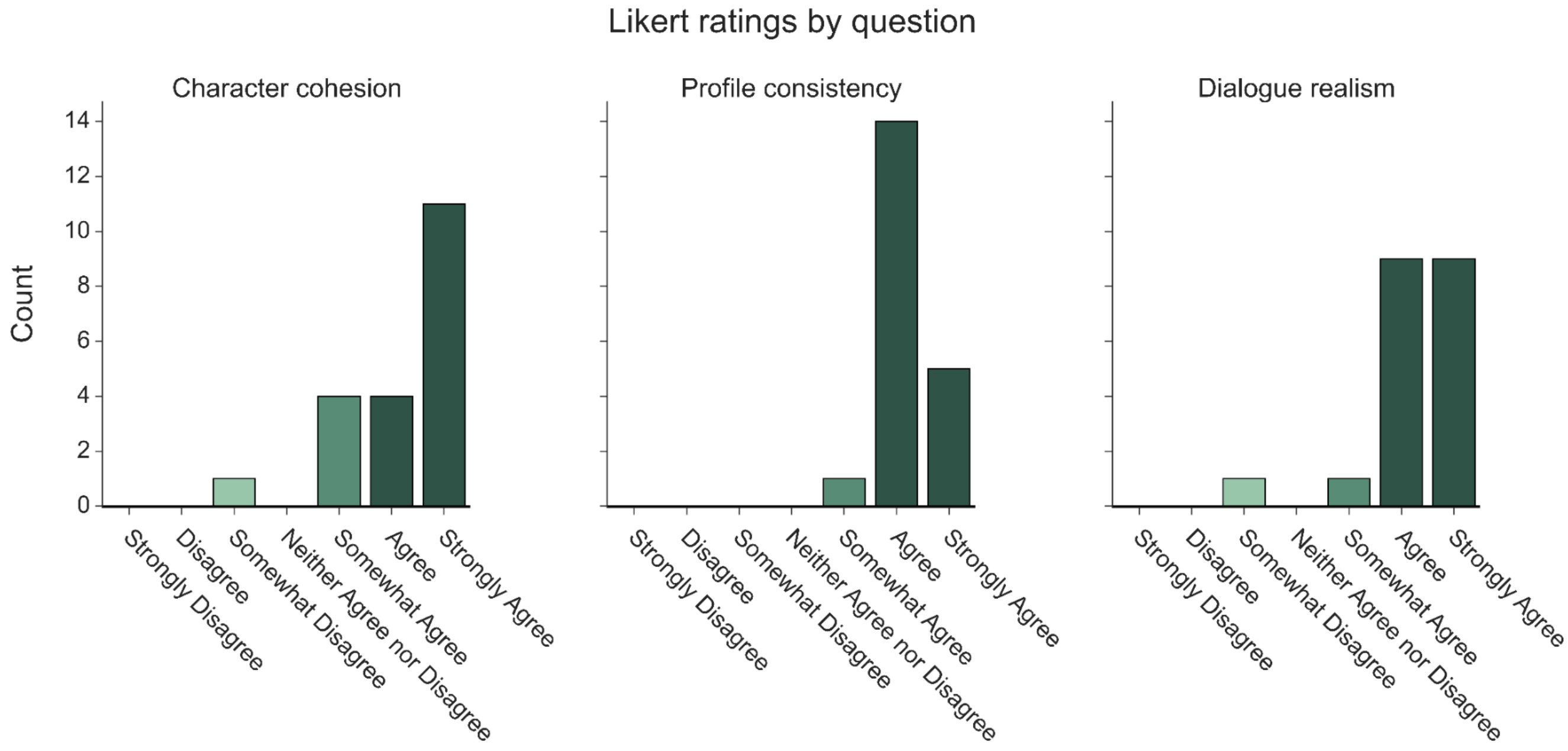

**Figure 6.** Likert-scale responses to the three qualitative questions across expert raters and interviews.

## Discussion

The results of this pilot study suggest that a voice-enabled virtual patient system based on a LLM can produce cohesive and clinically plausible simulations of interview-based clinical assessments. The primary findings show that experienced clinicians scored a set of virtual patients with high fidelity to their pre-defined symptom profiles and that inter-rater reliability point estimates were promising. These results provide initial evidence that LLM-based simulations can serve as a standardized and scalable tool to address critical gaps in clinical interview training, particularly for CNS clinical trials.

A notable finding in our study was the slight but systematic tendency for expert raters to score the virtual patients' symptoms as moderately more severe than their predefined parameters. This may suggest that the LLM, in its current configuration, produces richer or more evocative symptom descriptions than what is minimally required to meet a given scoring threshold, which could lead raters to "round up" when making a judgment. This presents a clear avenue for model refinement, possibly through modifications to the system's prompting or by fine-tuning the model on a corpus of expert-scored interview transcripts to more precisely capture the subtle nuances of symptom expression at each severity level.

Our quantitative analysis revealed a high point estimate for the inter-rater reliability for the total MADRS score (0.90) which is a promising, albeit preliminary, indicator of the system's ability to portray symptoms consistently. This estimate is comparable to the total ICC of 0.93 reported in the original validation of the structured interview guide (SIGMA) with human patients [29]. Our point estimates for the item-level ICCs are also strong, but these values should be interpreted with caution since the confidence intervals are wide, given our small sample of raters. The variability in item-level reliability also mirrors findings in the SIGMA study, which characterized its item-level reliability as "good to excellent." This suggests that such variability may reflect not just system-specific behavior, but also the inherent ambiguity in rating certain symptoms. Further evaluations with a larger sample of raters will be needed to assess this question.

The broadly positive qualitative feedback is an encouraging sign of the system's fidelity, although these findings should also be considered preliminary as the total number of interviews (N=20) is moderate. The wide agreement on profile consistency is a key finding. It suggests that even when experts' blind quantitative scores differed from the ground truth, they still perceived the patient's behavior as a reasonable representation of the intended clinical profile, indicating that the system's generated behavior falls within an acceptable spectrum of clinical interpretation.

This central finding is supported by the strong ratings for character cohesion (19/20 positive) and dialogue realism (19/20 positive). The isolated 'Somewhat Disagree' ratings are valuable as they point to specific areas for refinement. For instance, one rater's observation that despite factual consistency, the virtual patient appeared, at times, behaviorally inconsistent, highlights the challenge of maintaining nuanced, implicit manifestations of the symptoms throughout the entire conversation. Another comment suggested that the instances of unrealistic verbosity occurred specifically within the context of severe symptoms. This feedback directly suggests a need to calibrate the model's output to better match this typical clinical presentation. However, these individual observations did not appear to detract from the overall expert agreement. Taken together, these results provide initial evidence that the tool is capable of producing interactions that are clinically consistent, factually and behaviorally cohesive and linguistically natural.

These findings address long-standing limitations of current training paradigms. The system presented here offers a compelling alternative that combines the realism of SPs with the scalability and standardization afforded by automated solutions. By providing on-demand access to a wide range of patient personas, the virtual patient simulations could offer trainees the repeated practice opportunities needed to build expertise. Furthermore, the system's ability to generate consistent patient stimuli could be leveraged to calibrate raters and mitigate "rater drift" throughout the course of a clinical trial, contributing to enhanced data quality.

### Limitations and future directions

Despite the promising results, this study has several limitations that highlight important directions for future research. The most critical limitation is the study's sample size (4 virtual patient personas and 5 expert raters), which is appropriate for a pilot study but limits the broader generalizability of our findings. A larger-scale evaluation with more diverse patient characters and a larger set of raters is needed to validate the initial results from this pilot study.

Moreover, the current simulations are based only on audio, lacking any non-verbal cues such as facial expressions and body language. These cues are an essential part of clinical assessment, for instance in evaluating items such as "Apparent Sadness" on the MADRS, and their absence limits the system's ecological validity. An important next step is to integrate the voice pipeline into a multimodal simulation, for instance with animated 3-dimensional avatars, that can train raters to interpret both verbal and non-verbal communication.

While the virtual patients in our simulations demonstrated high narrative cohesion, it is important to acknowledge that LLMs are inherently susceptible to generating factually inconsistent or clinically implausible content. Even though this was not an issue in our controlled study, robust guardrails and continuous validation will be essential for deploying such a system in a real-world training environment.

The present study focused on symptoms of depression as measured specifically by the MADRS. In practice, this tool holds promise for supporting clinical assessment training across a broad range of psychiatric conditions and assessment scales such as for anxiety and bipolar disorder. Careful validation will be necessary to demonstrate the system's generalizability to these scenarios.

Finally, a critical future direction will be to rigorously test the system's pedagogical effectiveness in a real training scenario. While expert feedback suggests the tool would be beneficial, a longitudinal study is required to empirically determine if training with this simulation system can lead to demonstrable improvements in trainee scoring accuracy, confidence and interviewing skills compared to traditional training methods.

*Conclusions*

LLM-powered virtual patient simulations represent a viable novel tool for training clinicians in standardized clinical assessments. This pilot study provides preliminary evidence that the system can produce clinically plausible practice scenarios that experts can score with a high degree of accuracy and promising reliability estimates. By addressing the need for scalable, standardized and realistic training opportunities, this technology has the potential to enhance the quality of psychiatric assessments, improve data integrity in mental health clinical trials, and ultimately better prepare clinicians for real-world patient interactions. Future work will focus on expanding the system's capabilities and rigorously testing its impact on trainee performance over time to solidify its role in medical education.

## Acknowledgements

We thank the expert raters who participated in this study. Funding was provided by Brooklyn Health.

## Data Availability

User study data and analysis code are available upon request.

## Conflicts of Interest

None declared.

---

## Abbreviations

AI: artificial intelligence

CNS: central nervous system

ICC: intraclass correlation coefficient

LLM: large language model

MADRS: Montgomery-Asberg Depression Rating Scale

STT: speech-to-text

SP: standardized patient

TTS: text-to-speech

VP: virtual patient